# Complex structure of the carbon arc discharge for synthesis of nanotubes

V Vekselman, M Feurer, T Huang, B Stratton and Y Raitses

[1] Princeton Plasma Physics Laboratory, Princeton University, P.O. Box 451, Princeton, NJ 08543, USA

**Abstract**

Comprehensive non-invasive spectroscopic techniques and electrical measurements of the carbon arc revealed two distinguishable plasma synthesis regions in the radial direction normal to the arc axis. These regions, which are defined as the arc core and the arc periphery, are shown to have very different compositions of carbon species with different densities and temperatures. The colder arc periphery is dominated by carbon diatomic molecules ($C_2$), which are in the minority in the composition of the hot arc core. These differences are due to a highly non-uniform distribution of the arc current, which is mainly conducted through the arc core populated with carbon atoms and ions. Therefore, the ablation of the graphite anode is governed by the arc core, while the formation of carbon molecules occurs in the colder arc periphery. This result is consistent with previous predictions that the plasma environment in the arc periphery is suitable for synthesis of carbon nanotubes.

1. Introduction

Research into carbon based nanomaterials such as fullerenes, single-, and multi-wall nanotubes, nanofibers, and graphene has been driven by the unique physical properties of these nanomaterials and their potential applications, including but not limited to electronics, pharmaceuticals, and structural materials. Ablation of graphite electrodes by an arc discharge [1,2] and laser vaporization of graphite targets [3,4] are well established methods for a potentially large-scale production of these nanostructures. The DC arc discharge method has the advantages of being less expensive and easier to implement than laser vaporization techniques. For synthesis of carbon nanotubes (CNTs), a DC arc discharge with a consumed graphite anode at atmospheric pressure helium gas is commonly used [5-7]. Typical parameters of this carbon arc are: arc current of 50-100 A, arc voltage of 10-20 V, and helium gas pressure of 500-600 Torr. From visible observations, the typical diameter of the carbon arc for synthesis of nanomaterials does not exceed 1 cm, while the length of the arc measured as the distance between the anode electrode and the cathode electrode is approximately 0.1-0.2 cm.

When metal catalysts (e.g. Nickel and Yttrium) are added to the ablating graphite anode, the carbon arc can produce single-walled CNTs [5,8,9]. Without catalysts, the carbon arc produces multi-walled CNTs,

**Complex structure of the carbon arc discharge for synthesis of nanotubes**

which are usually found in a carbonaceous deposit formed on the cathode electrode [10]. In both cases, the graphite anode serves as a carbon feedstock for synthesis processes. The ablation of the graphite anode determines the synthesis yield of CNTs. For a large synthesis yield, the arc must operate in the so-called high ablation mode with current densities larger than 100 A/cm$^2$ [11]. However, arc operation in the high ablation mode is also accompanied by extensive formation of non-desired carbon byproducts such as soot particles [12]. This poor selectivity is the primary drawback of arc synthesis of nanomaterials and further understanding of the synthesis process is required to improve it.

Several experimental studies have attempted to develop improved control of the arc synthesis process with limited success [13,14]. This is in large part due to the fact that it is not known where and how the nanotubes are synthesized in the arc. Existing models of carbon arcs are not able to answer these questions because they do not include a self-consistent description of the arc plasma and synthesis processes [15-17]. Nevertheless, in Ref. [16], fluid simulations of the carbon arc coupled with reduced macroscopic models of the nanotube growth predicted a non-uniform distribution of the plasma temperature and the plasma density, both of which reach their maximum values at the center of the arc. The latter also implies a non-uniform distribution of the arc current between the arc axis and the arc periphery. This result is in qualitative agreement with measurements of the arc current distribution at the cathode using a segmented cathode [18]. Although the experiments showed a much greater non-uniformity of the arc current distribution than predicted by simulations, the arc periphery was found to have plasma environments (e.g. plasma temperature of less than 0.2 eV [16]) more suitable for the growth of CNTs than the hotter arc core (temperature of ~ 0.5-1 eV) in which CNTs would be likely destroyed by sublimation. In more recent experiments with a fast movable extraction probe, single-walled CNTs were indeed found in the arc periphery [7], well away from the arc core.

What remains unknown is the composition of plasma species in different regions of the arc. This is because, with the exception of the production of carbon atoms and ions, the existing models do not include plasma-chemical reactions that occur in the arc. This is in contrast to previous optical emission spectroscopy (OES) measurements which showed that di-atomic carbon dominates the chemical composition of the carbon arc [19-24]. Moreover, the temperature of $C_2$ may not be the same as the temperature of other plasma species, including ions, electrons and other non-ionized species. This may be especially true in the arc periphery and in the near-electrode regions, in which plasma is expected to be non-equilibrium (see for example Refs. [25,26]). Substantial deviations from equilibrium can be induced by difference in ionization/recombination kinetics, electron/ion/rotational/vibrational temperatures, reaction rate coefficients and other local gradients of plasma properties (density, thermal conductivity, *etc*). Finally, recent experiments [27,28] demonstrated unstable behavior of the carbon arc, which was not

**Complex structure of the carbon arc discharge for synthesis of nanotubes**

captured by previous experimental studies and could not be predicted or explained by existing arc models limited to a steady state description. Addressing these issues is necessary for understanding of the synthesis processes in the arc and the development of predictive models.

In this paper, optical emission spectroscopy (OES), spectrally-resolved fast frame imaging (FFI) and planar laser induced fluorescence (PLIF) are applied to characterize the plasma composition, plasma density and temperature in the carbon arc. The most important new finding of this study is that the arc core and the arc periphery are very distinguishable regions of the arc with different chemical compositions, temperatures and densities of the plasma species. Based on these results, it is suggested that the arc core is responsible for the ablation of the graphite anode which provides the carbon feedstock to the arc plasma. We provide the first direct evidence that the ablation of the graphite anode is governed by the near-anode process in the arc core. This result confirms a hypothesis proposed by [11] and later modeled by [26] that the ablation rate is not only a function of the current density, but also depends on the anode voltage drop. Finally, we show the formation of carbon molecules takes place in the colder arc periphery. This paper is organized as follows. Section 2 describes the experimental setup and diagnostics. Experimental results are described and discussed in Section 3. Conclusions of this work are summarized in Section 4.

## 2. Experimental setup

The arc synthesis setup used in these experiments is described elsewhere [11,18,29,30]. The experiments reported here were conducted mainly with a 0.65 cm diameter graphite anode and a 0.95 cm diameter graphite cathode. This arc electrode configuration is referred to as C9-A6. The arc was powered by a current regulated power supply. Most of experiments were conducted with horizontally aligned electrodes which can be positioned with a high precision of ~10 μm with respect to the optical axis using two computer controlled sliders.

The electrical circuit of the arc includes a ballast resistor to mitigate relaxation oscillations associated with a negative differential resistance of the arc [31], a low resistance shunt in series with the arc discharge to measure the arc current, and a 1:10 voltage divider in parallel with the arc to measure the voltage drop across the arc and the arc electrodes. A National Instruments data acquisition card (PCI-6251) was used to digitize (at 1 kHz rate) measurements of the voltage and current waveforms of the arc discharge, gas pressure and positions of both electrodes.

The arc configuration and OES, FFI and PLIF diagnostics are shown in Fig 1. The optical axes of the OES, FFI and PLIF diagnostics were fixed in space and intersected the middle of the axis between the electrodes at right angles. The OES setup is based on an imaging spectrograph (Chromex is250) coupled

**Complex structure of the carbon arc discharge for synthesis of nanotubes**

with a CCD camera (FLIR FL3) and a fiber-coupled low-resolution spectrometer with linear CCD (Ocean Optics 2000+). Spectral images were acquired by imaging the plasma on the entrance slit of the spectrograph with the radial direction in the arc discharge oriented along the long dimension of the slit. The use of collecting optics with a magnification of M=1/4 resulted in spatial resolution of 120 pix/mm along the slit. The linear dispersion of the spectral images in direction perpendicular to the slit was 0.0046 nm/pixel at a wavelength of 656.28 nm and entrance slit width of 20 μm.

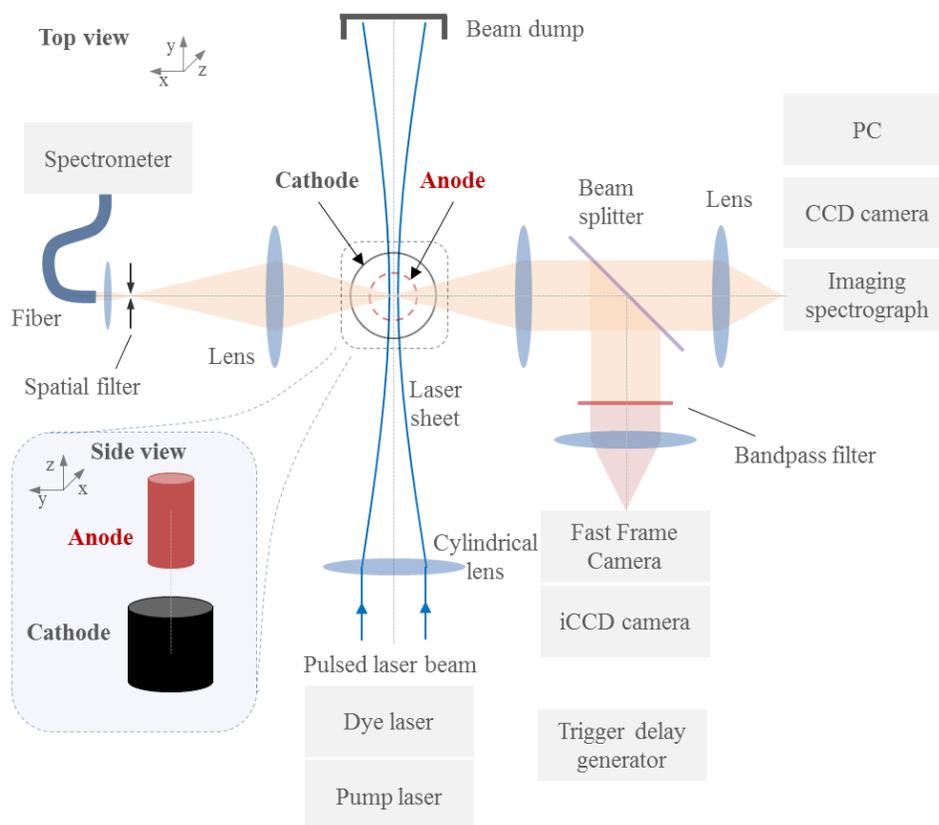

**Figure 1**. Schematic of the experimental setup of carbon arc for synthesis of carbon nanotubes.

The low-resolution spectrometer (dispersion of 0.22 nm/pix) was used to acquire spectra in the visible range of 380-700 nm. The FFI system consists of Phantom V7.3 camera coupled with a macro objective lens and a set of bandpass interference filters, see Table 1.

Table 1. List of FFI diagnostics parameters.

| Model | Center wavelength | Bandwidth | Plasma species |
|---|---|---|---|
| Andover 470FS10-50 | 470 nm | 10 nm | $C_2$ |
| Thorlabs FB600-10 | 600 nm | 10 nm | C I |
| Andover 656FS02-50 | 656.3 nm | 1 nm | H |

**Complex structure of the carbon arc discharge for synthesis of nanotubes**

The PLIF setup consists of a light source for planar illumination across the arc and acquisition system based on iCCD camera (Andor iStar 734) coupled with bandpass filter Andover 470FS10-50, see Table 1. A dye laser (Spectra-Physics Cobra-Stretch-G) pumped by Nd:YAG laser (Continuum SL III-10, 8 ns) was used as a light source to excite (3,1) Swan band at 437.14 nm with fluorescence near 469 nm. Laser beam shaping to form laser sheet of 0.6 cm height was done by a cylindrical lens Thorlabs LJ4147. The laser pulse energy was measured by Thorlabs PM100D energy meter coupled with ES220C pyroelectric sensor.

The arc experiments were conducted in a helium-hydrogen gas mixture (ratio 95/5 by weight) at a background pressure of 67 kPa (500 Torr). A small fraction of hydrogen was added to facilitate spectroscopic characterization of the arc plasma based on the hydrogen Balmer lines ($H_\alpha$ and $H_\beta$). It was verified that this small addition of hydrogen did not affect the arc operation, including voltage and current of the arc discharge, ablation and deposition rates.

To initiate the arc, the electrodes were first brought into contact and then separated to produce a self-sustained discharge with the desired gap between the electrodes. In contrast to previous studies on the same arc setup [11,18,29,32], the inter-electrode gap length (~ 0.1-0.2 cm) was monitored by a digital camera during the arc operation. Change in the gap length due to anode ablation or deposit growth on the cathode was compensated for by adjustment of the positions of the electrodes. This auxiliary manual control of the gap was implemented in addition to digital feedback control of the gap as the function of the arc voltage, which is commonly used in studies of carbon arcs for synthesis of CNTs [11,15,18,29,32]. We found it critically important to monitor the gap via the digital camera in order to obtain consistent and reproducible results, especially for the arc operation with high ablation rates.

The arc current was varied to operate the arc in either low ablation mode or high ablation mode. These modes are similar to the ablation modes reported in Ref. [11,32] for a given current with different anode diameters. However, for spectroscopic measurements, the variation of the arc current is more convenient because it provides the real-time control of the arc operation mode. In the low ablation mode, the ablation rate of the graphite anode is nearly constant and independent of the arc current. In the high ablation mode, the ablation rate increases rapidly with increasing current, as shown in Fig. 2. Here ablation and deposition rates were measured by electrodes weighting after arc run of 60 s. For a 0.65 mm diameter anode, the transition between the low ablation mode and the high ablation mode occurs at 55 A.

**Complex structure of the carbon arc discharge for synthesis of nanotubes**

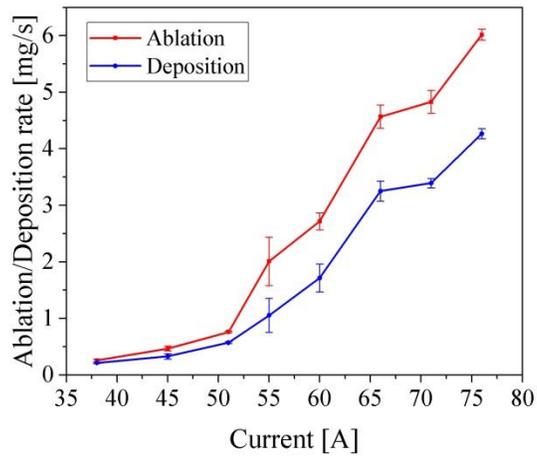

**Figure 2**. Ablation and deposition rates as a function of the discharge current. Electrode configuration C9-A6, helium gas at 67 kPa.

3. **Results and discussions**

   *3.1. Arc composition*

A typical visible spectrum of the arc radiation obtained using the low-resolution spectrometer is shown in Fig. 3.

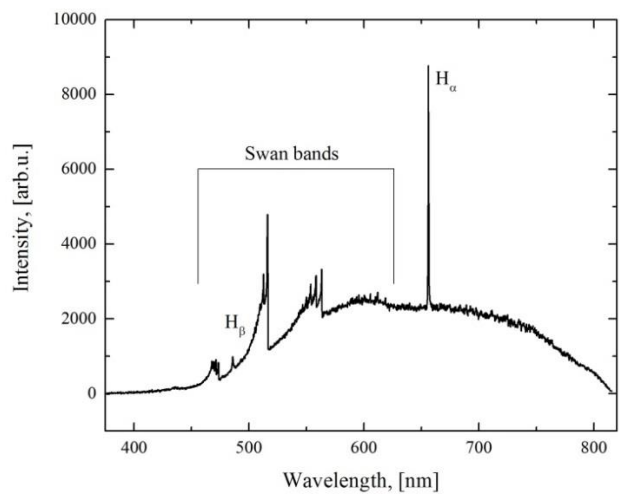

**Figure 3**. Typical low-resolution arc spectrum.

**Complex structure of the carbon arc discharge for synthesis of nanotubes**

The discrete spectrum of atomic and molecular lines is superimposed on a broadband continuum due to emission directly from the arc electrodes. As in previous OES studies of carbon arcs (for example, Refs. [[21,23,33-36]), spectral lines of atomic carbon neutrals and ions were detected but the most intense features are the $C_2$ Swan bands. Thus, the arc is composed mostly of carbon species produced by the ablation of the graphite anode. Due to the low excitation rate for helium atoms (mostly by stepwise excitation), the helium spectral lines are significantly weaker than spectral lines of all carbon species. This result also implies a lack of high energy electrons which could excite helium atoms. For comparison, the population ratio of electrons with energies 12.08 eV ($n=3$ hydrogen state) and 21.21 eV (first excited state of helium) at $T_e=1$ eV temperature is $10^4$. Thus, it is unlikely that helium gas is directly involved in plasma generation, although it may still affect the plasma through heat transfer in the arc and between the arc and the surrounding environment, as well as through convection [35].

*3.2. Arc structure*

Spectrally resolved FFI images obtained at 656 nm (hydrogen $H_\alpha$ line), 600 nm (C I), and 470 nm ($C_2$ Swan band) with the arc operated in the low ablation mode are shown in Fig. 4.

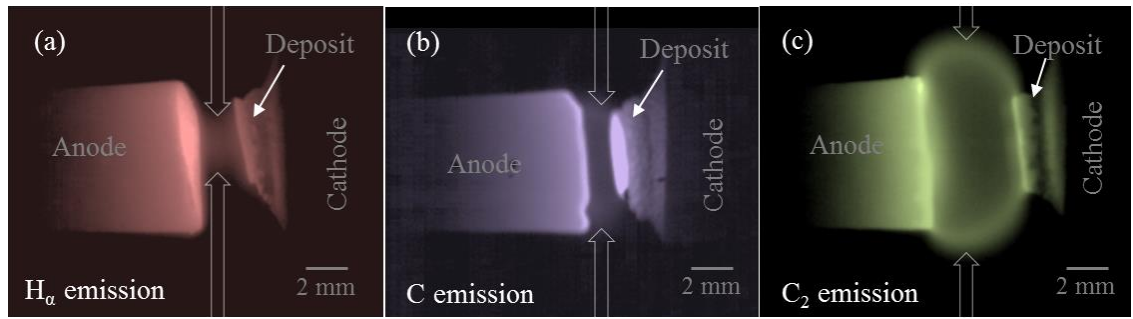

**Figure 4**. Spectral images of $H_\alpha$ (a), C I (b), and $C_2$ (c) obtained by FFI with the arc operated in the low ablation mode (graphite electrodes, configuration C9-A6, 50 A). Intensity and colors were artificially adjusted to enhance visualization of the emission patterns.

These images represent the line-integrated radiation intensity distribution of selected plasma species in the arc; the corresponding excitation energies are 12 eV ($H_\alpha$), 10.7 eV (C I) and 2.6 eV ($C_2$), respectively. White arrows denote the boundary of the emission patterns. The $H_\alpha$ emission region is constricted toward the electrode axis, the C I emission extends farther in the radial direction, and $C_2$ emission covers all radial distances and exceeds the anode radius. The cross-sectional radiation distribution of selected species taken in the middle of the inter-electrode gap is shown in Fig. 5a. The profiles are normalized to the peak intensity of the $C_2$ radiation but without correction for variation of camera sensitivity versus wavelength.

**Complex structure of the carbon arc discharge for synthesis of nanotubes**

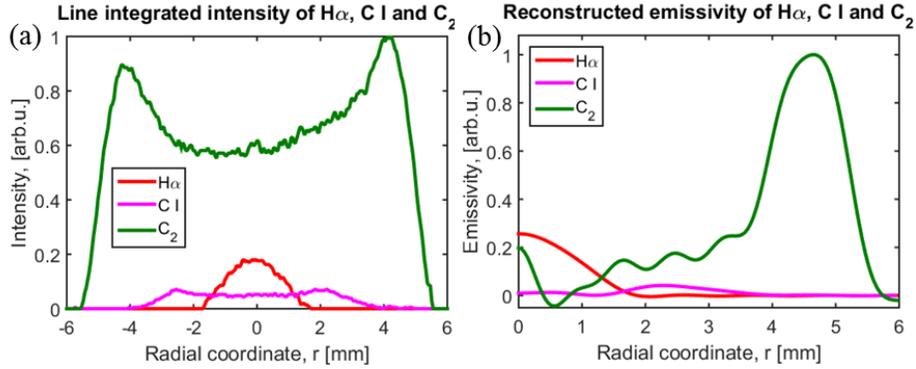

**Figure 5**. (a) Radial profiles of line-integrated radiation intensity from Fig. 3 taken in the middle of the inter-electrode gap, (b) reconstruction of emissivity for each selected species in (a) for positive radial coordinates.

The sequence of radiation intensity peaks from the arc center ($r = 0$ cm) follows decrease of corresponding upper energy level of the radiative transitions (Fig. 5). Since these measurements represent line-of-sight integrated intensities, an inversion algorithm must be applied to reconstruct actual emission coefficients. Assuming rotationally symmetric radiation, an Abel inversion was used to reconstruct the radial distribution of emission coefficients shown in Fig. 5b from the line-integrated intensities of Fig. 5a.

The region of $H_\alpha$ emission corresponds to the highest electron temperature (assuming hydrogen is distributed uniformly in the chamber) for excitation and is associated with the arc core, the hot, dense region where most of discharge current is conducted [25]. Its diameter of ~ 0.4 cm is consistent with the arc current measurements with segmented cathode [18], which showed that about 70% of the arc current to the cathode is conducted through a 3.2 mm diameter region at the cathode surface. The emission coefficients of C I and $C_2$ have a hollow profile with a clearly defined boundary layer and decrease towards the arc core. The lower C I emission in the arc core region is probably due to ionization of carbon atoms and their resulting depletion in this region. The sensitivity of the Phantom camera was not adequate to detect carbon ion lines, due to the low excitation rate. The lowest excited level of C II with emission in visible wavelength range is at 16.3 eV.

The $C_2$ fluorescence images was obtained by planar LIF and shown in Fig. 6. The laser beam was attenuated to 0.1 mJ per pulse in order to avoid saturation effects. Since the $C_2$ Swan bands involve the lower electronic state $^3\Pi_u$, which is ~0.18 eV above the ground level [37], this LIF excitation scheme accurately represents the $C_2$ distribution.

**Complex structure of the carbon arc discharge for synthesis of nanotubes**

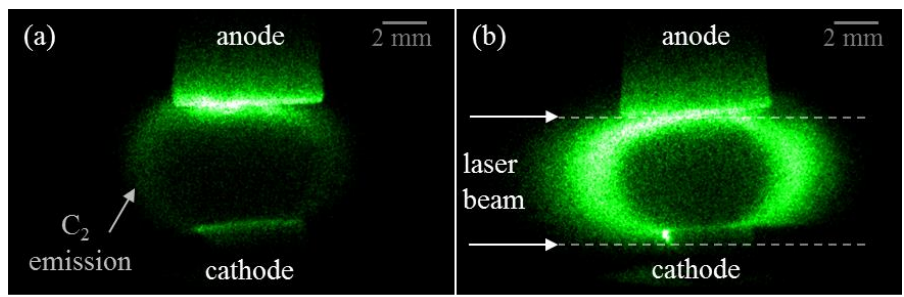

**Figure 6**. Emission pattern of $C_2$ with laser (a) off and (b) on (PLIF). Graphite electrodes, configuration C9-A6, 50 A, exposure time 20 ns.

The Fig. 6a shows the $C_2$ emission acquired during a 20 ns exposure time without the laser pulse. The emission pattern has a typical hollow profile similar to Fig. 4c. The PLIF image shown in Fig. 6b confirms the depletion of $C_2$ in low excited states from the central arc region. Appearance of a narrow $C_2$ emission region near the anode surface in Fig 6a and 6b is indicative of the ejection of $C_2$ from the anode material. Since the bond dissociation energy of $C_2$ is 6.3 eV, highly excited carbon dimers may exist in the arc core. However, these results show that the primary formation of carbon diatomic structures (by carbon condensation) occurs in the arc periphery region.

The observed non-monotonic distribution of carbon species is indicative of specific conditions in the arc discharge. Similar to the emission rate coefficients which are a function of the emitter density and temperature, the nanostructure growth rate depends on the flux of feedstock material and the local temperature. This result may explain non-monotonic distribution of the nanoparticles which was recently reported in Ref. [7].

In the high ablation mode, the arc behavior can be highly unstable, resulting in oscillations of the discharge current and the radiation pattern due to the arc motion [28]. However, even during this arc motion with a characteristic speed of tens m/s, the layered structure of the arc is generally preserved, as shown in Fig. 7.

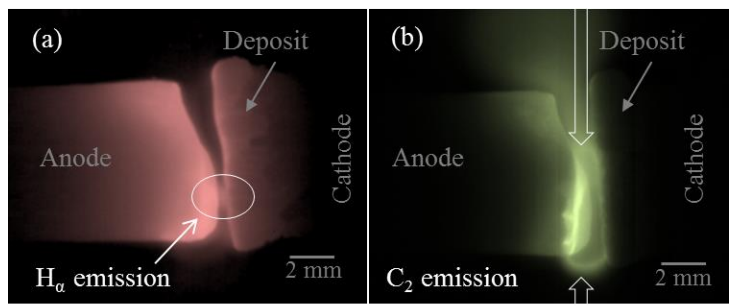

**Complex structure of the carbon arc discharge for synthesis of nanotubes**

**Figure 7**. Simultaneous frames of the arc $H_\alpha$ (a) and $C_2$ (b) emission in high ablation mode (graphite electrodes, configuration C9-A6, 65 A, exposure time 2 μs). Intensity and colors were artificially adjusted to enhance visualization of the emission patterns.

Figs. 7a and 7b show single FFI frames for $H_\alpha$ emission and $C_2$ emission, which were acquired simultaneously during arc operation. These images show the same structural pattern of the arc as the pattern obtained for the low ablation mode (Fig. 4). In particular, the arc core region is surrounded by $C_2$ emission with radial symmetry and a characteristic hollow profile. As a result of the arc motion in the high ablation mode, remnants of $C_2$ emission can be observed (Fig. 7b) that reveal a complex time-varying density $C_2$ distribution in the arc discharge.

*3.3. Arc core*

To confirm the coincidence of the observed $H_\alpha$ emission pattern with the arc core region detailed measurements of the electron temperature and density were obtained from the hydrogen spectra measured by OES. In OES the self-absorption effect can significantly influence on accuracy of quantitative analysis of spectroscopic data. Therefore the degree of self-absorption of hydrogen Balmer lines was checked by doubling the emission from the plasma using a spherical mirror behind it. This procedure confirmed that plasma is optically thin for spectral lines of interest.

The Stark broadening of hydrogen spectral lines and the Boltzmann diagram method were implemented to obtain density and temperature of the plasma [38,39]. Spectral images of the Balmer $H_\alpha$ line were processed to obtain the Stark broadening component by deconvolution of the Voigt profile of measured spectral line shapes (see detailed description of deconvolution procedure in Ref. [40]). The choice of $H_\alpha$ over $H_\beta$ was determined by better SNR and more uniform background (the spectral region of $H_\beta$ was populated by molecular bands of $C_2$). A typical spectral image of the $H_\alpha$ line where the vertical axis is the spatial coordinate is shown in Fig. 8a.

**Complex structure of the carbon arc discharge for synthesis of nanotubes**

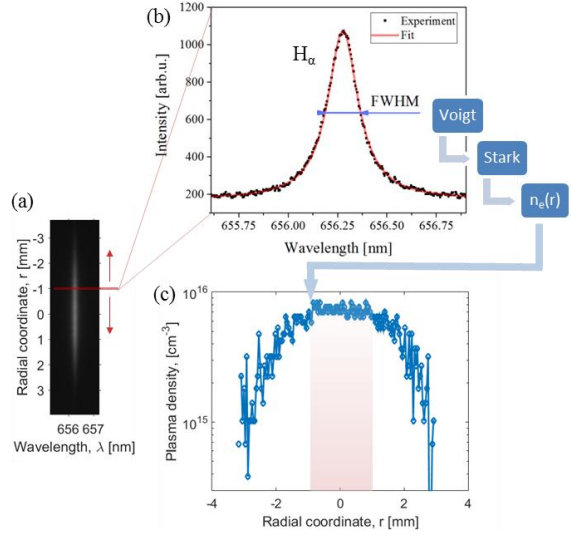

**Figure 8**. Schematic of radial plasma density calculation using spectral image of $H_\alpha$ (a) from the middle of the inter-electrode gap: horizontal slice of spectral image reveals spectral line profile (b) which is further de-convolved to obtain Stark broadening and $n_e(r)$ at given radial coordinate. The result is a radial density profile, $n_e(r)$ (c). (Low ablation mode, 21 V, 55 A, C9-A6).

Each row of the obtained spectral image was transformed to a spectral line profile (Voigt profile), Fig. 8b, and then de-convolved to obtain the full width at half maximum (FWHM) of the Stark component, $\Delta\lambda^{Stark}$

$$\begin{cases} V(\lambda) = A\int \exp\left(-\frac{4\ln 2(\lambda'-\lambda_0)^2}{G_w^2}\right) \frac{L_w^2}{4(|\lambda-\lambda'|-\lambda_0)^2 + L_w^2} d\lambda' \\ L_w = \Delta\lambda^{Stark} + \Delta\lambda^{WV} \\ G_w = \sqrt{\left(\Delta\lambda^{Doppler}\right)^2 + \left(\Delta\lambda^{Instr}\right)^2} \end{cases} \quad (1)$$

where $V(\lambda)$ is a Voigt function, $G(\lambda)$ and $L(\lambda)$ are Gaussian and Lorentzian functions with FWHM $G_w$ and $L_w$, correspondingly. $\Delta\lambda^{Instr}$ is the instrumental broadening FWHM measured with a spectral calibration lamp, and $\Delta\lambda^{Doppler}$ and $\Delta\lambda^{WV}$ are Doppler and Van-der-Walls broadening FWHMs. The temperature obtained by the Boltzmann diagram method was used to estimate the Doppler broadening. The Stark FWHMs were then transformed into the radial electron density profile using data of Gigosos, et.al. [41,42] as shown in Fig. 8c. The central region of the density profile is flat with a density of $8\cdot10^{15}$ cm$^{-3}$ inside ~0.2 cm (shaded region in Fig. 8c), which is identified as the arc core. The density drops to

**Complex structure of the carbon arc discharge for synthesis of nanotubes**

$8 \cdot 10^{14}$ cm$^{-3}$ at the edges. A similar procedure was applied to analyze the arc discharge plasma in high ablation mode and yielded a plasma density up to $3 \cdot 10^{16}$ cm$^{-3}$ in the center part of the arc core (Fig. 9). The measured plasma densities increase monotonically with discharge current.

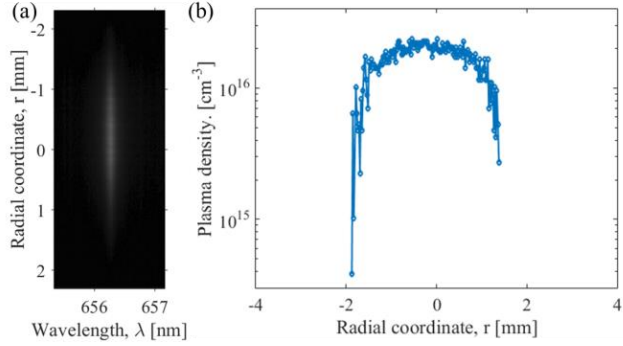

**Figure 9**. (a) Spectral image of H$_\alpha$ used to calculate (b) radial density profile, $n_e(r)$, in high ablation mode (20 V, 65 A, C9-A6).

The electron temperature was obtained using the Boltzmann diagram method. It relies on the connection between the populations of the radiative states and the temperature of the perturbers responsible for excitation. However, this method cannot be directly applied to the arc plasma because it may deviate from the LTE state [43,44], as was confirmed by data from the ADAS database [45]. Therefore, we applied a collisional radiative model to obtain the population ratio of spectral lines of the Balmer series, H$_\alpha$, H$_\beta$ and H$_\gamma$. The fiber-coupled low resolution spectrometer was used to collect spectra from the middle of the arc inter-electrode gap with 1 μs exposure time and an observational cone cross-sectional area in the focal plane of ~0.8 mm$^2$. The spectral sensitivity of the setup was calibrated with a black body source (Labsphere SC6000). The electron temperature for each set of Balmer lines (*m-n*) was calculated from

$$kT_e = -\frac{E_m - E_n}{\log\left(\dfrac{k_{sys}^{mn}}{K_{mn}} \dfrac{I_{m-2}^{meas}}{I_{n-2}^{meas}}\right)} \quad (2)$$

where *m* and *n* are the quantum numbers of the relevant hydrogen states, $I_{n-2}^{meas}$ ($I_{m-2}^{meas}$) is the measured intensity of the spectral line corresponding to the transition between the $E_n$ ($E_m$) and $E_2$ energy levels,

$$K_{mn} = \chi_{mn} \frac{\lambda_{n-2} \cdot \sum g_m A_{m-2}}{\lambda_{m-2} \cdot \sum g_n A_{n-2}} \quad (3)$$

**Complex structure of the carbon arc discharge for synthesis of nanotubes**

is a coefficient accounting for the degeneracy of levels (*g*), spontaneous emission coefficients (*A*), transition wavelengths ($\lambda$) and CRM correction factor $\chi_{mn}$. Here $k_{sys}^{mn}$ is a coefficient of the spectral response function of the optics and spectrometer calculated using the calibrated black body source. For the low ablation mode, the time evolution of the electron temperature during an arc run is shown in Fig. 10.

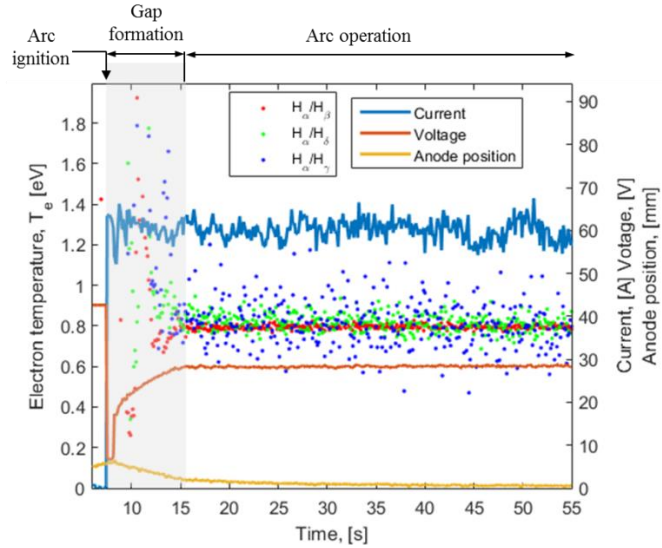

**Figure 10**. Time evolution of the arc core plasma electron temperature during an arc run. The temperature (dotted plot) was calculated from ratios of $H_\alpha/H_\beta$ (red), $H_\alpha/H_\gamma$ (green) and $H_\alpha/H_\delta$ (blue). Solid lines are waveforms of discharge current (blue), voltage (dark red) and axial position of the anode (yellow).

In Fig. 10, region of the arc ignition and formation of inter-electrode gap is shaded since it does not provide a reliable temperature measurement. Once the gap is formed the measured electron temperature is 0.8±0.1 eV and it remains constant during the arc operation. In high ablation mode the temperature behaves similarly and reaches 0.9±0.2 eV. In the high ablation mode case, smaller gaps and intense formation of particles lead to noisier data and therefore larger uncertainty in calculated temperature.

Summarizing this section, the plasma density and the electron temperature in the arc core exhibit a gradual increase with the discharge current. Thus plasma in the inter-electrode space is unlikely responsible for rapid change in ablation rate during transition from low ablation mode to high ablation mode.

### 3.4. Anode processes

In Refs. [11,25,46], it was proposed that for the high ablation mode, the power deposition at the anode increases due to the increase of the anode voltage drop. The validation of these predictions is difficult

**Complex structure of the carbon arc discharge for synthesis of nanotubes**

because the anode voltage drop develops along a very short near-anode region of tens of microns [47]. Previous measurements of the arc Volt-Ampere characteristics at large gaps were not reveal predicted non-monotonic increase of arc voltage [11,48]. However, non-linear dependence of arc voltage on the gap length was observed for small gaps (<2 mm) [48]. Measurements of the space potential or electric field with, for example, electrostatic probes in such a small plasma region are not practical. Therefore, we applied a different approach to determine the anode voltage drop by measuring the arc voltage at a given arc current at different inter-electrode gaps [49]. In particular, changes of the arc voltage were monitored during the initiation of the arc discharge from the moment when the electrodes are brought into contact and then separated until the arc reaches a quasi-steady state condition (Fig. 11). The arc voltage was also monitored during the extinction of the arc, which was accomplished by bringing the electrodes back into contact, i.e. shorted. During the initiation and the extinction processes, the anode electrode was moved with a constant speed.

When the electrodes are brought in contact (Fig. 11 a-b), the current starts to flow and the voltage drops to $V_{el} = IR_{el}$ (Fig. 11 b-c) where $I$ is the discharge current and $R_{el}$ is the resistance of the graphite electrodes. After the electrodes are separated the arc discharge is initiated (Fig. 11 c-d) if the cathode surface temperature is sufficient to sustain the arc current by thermionic emission.

Development of the arc is accompanied by the formation of cathode and anode regions with corresponding voltage drops and a plasma positive column which connects these regions. At the initiation of the arc discharge there is no space for plasma column formation and the measured arc voltage is the total of the cathode and anode voltage drops [31]. The anode voltage drop can be assumed to be small compared to the cathode voltage drop. This is a typical situation for gas discharges [31]. This assumption also implies that the anode ablation is not strong during this initial phase of the arc. This has been confirmed in experiments [50]. Once the cathode and anode regions are formed the arc column starts to develop, increasing the arc voltage due to length-dependent energy losses, including but not limited to electron collisions with heavy atomic and molecular species in the plasma. A relatively high conductivity of the arc column results in a steeper and almost constant voltage increase as a function of the gap length. Hence, the sum of the near-electrodes voltage drops during the arc initiation, $V_c + V_a^{in}$, where $V_c$ is the cathode voltage drop and $V_a^{in}$ is the anode voltage drop during the initiation of the arc, can be found from the measured voltage waveform as shown in Fig. 11.

**Complex structure of the carbon arc discharge for synthesis of nanotubes**

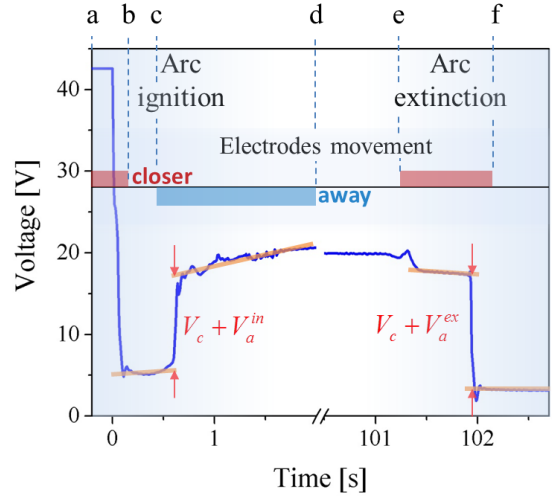

Fig. 11. Description of the anode fall and cathode fall measurements. Discharge voltage waveform (blue line) shown during arc ignition (a-d) and extinction (e-f). Electrodes are moving towards each other during (a-b) and (e-f) and outwards during (c-d). Red arrows indicate measurement points.

We shall now apply the same analysis of the near-electrode voltage drops to the extinction of the carbon arc. The main difference in the extinction of the carbon arc from its initiation is that the anode is already hot and the arc operates in the so-called anodic mode with a higher ablation of the graphite material than at the arc initiation. The sum of the near-electrode voltage drops, $V_c + V_a^{ex}$, can be obtained by eliminating the arc column region during shorting of the electrodes (Fig. 11 e-f). Here, $V_a^{ex}$ is the anode voltage drop measured during the arc extinction. Since the arc current at the cathode is maintained mainly by thermionic emission [29], we may assume the same cathode voltage drop for the initiation and the extinction of the arc. Hence, the change in the anode voltage drop of the arc is

$$\Delta V_a = V_a^{ex} - V_a^{in} \qquad (4)$$

It should be noted here that Eq. (4) provides with the low limit of the anode voltage drop variation. As was shown in [18,25] the increase of the latent heat flux towards the cathode in high ablation mode diminishes the importance of the ion flux as a heating mechanism of the cathode. This may lead to a decrease of the cathode voltage drop. Therefore, the actual anode voltage drop may exceed the values obtained from Eq. (4) by corresponding decrease of the cathode voltage drop. Fig. 12 shows the variation of the anode voltage drop calculated using Eq. (4) as a function of discharge current. The anode voltage drop increases rapidly as the arc current increases above 55 A. Following Ref.[26], the heat flux from the plasma to the anode can be expressed as

**Complex structure of the carbon arc discharge for synthesis of nanotubes**

$$q_a^e = j_e \left(2.5 T_e + \phi_w + V_a^{in} + \Delta V_a \right) \tag{5}$$

where $j_e$ and $T_e$ are the electron current density and temperature, respectively, and $\phi_w$ is the work function of the anode material. Since the arc current does not cause significant changes of the electron temperature and changes in the area of the arc attachment to the anode, Eq. (4) implies that the anode voltage drop is the only term that can lead to a rapid increase of the heat flux to the anode. The increase of the heat flux can lead to the increase of the anode temperature and the ablation rate, which is determined by the anode temperature. Thus, the transition from the low ablation mode to the high ablation mode with the arc current and the increase of the ablation rate with the arc current in the high ablation rate is associated with the increase of the anode voltage drop. This result is similar to results reported for the carbon arc with different anodes [32] and modeled in Ref. [46]. In these previous studies, for a given arc current, the transition from the low ablation mode to the high ablation mode took place with the reduction of the anode diameter. The transition was also predicted due to changes in the anode voltage drop.

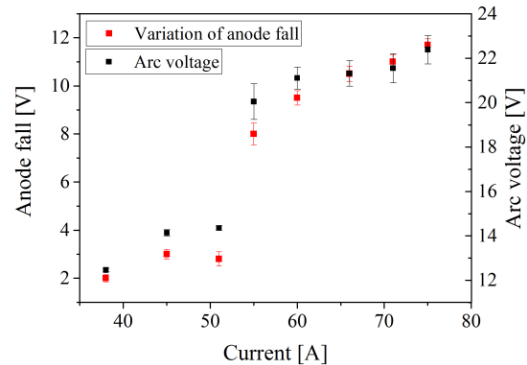

Figure 12. Variation of the anode voltage drop and arc voltage vs. the discharge current.

## 4. Conclusions

A carbon arc discharge for nanomaterial synthesis was comprehensively studied using non-invasive spectroscopic techniques and electrical measurements. Two observed arc plasma regions defined as the arc core and the arc periphery are shown to have very different compositions of carbon species with different densities and temperatures. These results suggest that these regions play also different roles in nanomaterial synthesis. The arc core represents the dense and hot plasma region conducting most of the discharge current. The latter is self-consistently sustained by processes in the cathode vicinity whereas formation of feedstock material for synthesis is governed by the anode ablation processes. Enhancement of the anode material ablation is achieved by increase of the anode voltage drop rather than increase of the current density. This result of direct electrical measurements of the voltage drop in the arc confirms

**Complex structure of the carbon arc discharge for synthesis of nanotubes**

hypothesis proposed by [11] and later modeled by [26] and shows importance of plasma surface processes in the synthesis arc. In contrast to the hot arc core, the arc periphery is colder and characterized by intensive formation of carbon molecules. The resulting radial distribution of carbon molecules has a distinguished hollow profile structure which is preserved regardless of the arc operation mode. This result corroborates previous theoretical predictions that plasma environments in the arc periphery are suited to synthesis of carbon nanotubes.

**Acknowledgements**

The authors wish to thank Dr. A. Gerakis, Dr. S. Gershman, Dr. S. Yatom and Dr. M. Shneider for fruitful discussions. We also thank Mr. Alex Merzhevskiy for his technical support of this work. This work was supported by U.S. Department of Energy, Office of Science, Basic Energy Sciences, Materials Sciences and Engineering Division.

**Complex structure of the carbon arc discharge for synthesis of nanotubes**

**Complex structure of the carbon arc discharge for synthesis of nanotubes**